\documentclass[CERN,manyauthors]{cernphprep}

\usepackage[comma,square,numbers,sort&compress]{natbib}
\usepackage[pdftitle={ESPP 2026: the NA60+/DiCE experiment},
        pdfauthor={NA60+/DiCE Collaboration},
        pdfsubject={},
        bookmarksopen=false]{hyperref}
\usepackage{lineno}
\usepackage{bm}
\usepackage{xcolor}
\usepackage{textcomp}
\usepackage{parskip}
\usepackage{enumitem}
\setlist{nolistsep}

\newcommand{\PbPb}{\ensuremath{\text{Pb--Pb}}\xspace}

\newcommand{\sqrtsNN}{\ensuremath{\sqrt{s_{\mathrm{\scriptscriptstyle NN}}}}\xspace}

\newcommand{\ccbar}{\ensuremath{\mathrm{c\overline{c}}}\xspace}

\newcommand{\Dmeson}[1]{\ensuremath{\mathrm{D}^{#1}}\xspace}

\newcommand{\Dzero}{\Dmeson{0}}
\newcommand{\Dplus}{\Dmeson{+}}

\newcommand{\Ds}{\ensuremath{\mathrm{D}^{+}_{\rm s}}\xspace}

\newcommand{\lambdac}{\ensuremath{\Lambda_{\rm c}}\xspace}

\newcommand{\lambdacplus}{\ensuremath{\Lambda_{\rm c}^+}\xspace}

\newcommand{\GeV}{\ensuremath{\text{~GeV}}\xspace}

\newcommand{\GeVcc}{\ensuremath{\text{~GeV}/c^2}\xspace}

\newcommand{\beq}{\begin{equation}}
\newcommand{\eeq}{\end{equation}}
\newcommand{\beqn}{\begin{eqnarray}}
\newcommand{\eeqn}{\end{eqnarray}}

\newcommand{\beqa}{\begin{eqnarray}}
\newcommand{\eeqa}{\end{eqnarray}}
\def\lsim{\raise0.3ex\hbox{$<$\kern-0.75em\raise-1.1ex\hbox{$\sim$}}}
\def\gsim{\raise0.3ex\hbox{$>$\kern-0.75em\raise-1.1ex\hbox{$\sim$}}}

\newcommand{\pt}{\ensuremath{p_{\mathrm{T}}}\xspace}

\newcommand{\pythia}{{\sc Pythia}\xspace}
\newcommand{\powheg}{{\sc Powheg}\xspace}

\newcommand{\vrtx}{vertex spectrometer\xspace} 

\begin{document}%

\begin{titlepage}
\title{European Strategy for Particle Physics 2026:\\ the NA60+/DiCE experiment at the SPS}
\ShortTitle{ESPP 2026: the NA60+/DiCE experiment}

\Collaboration{NA60+/DiCE Collaboration\thanks{See \hyperref[app:collab]{Appendix} for the list of collaboration members}}
\ShortAuthor{NA60+/DiCE Collaboration}
\bigskip
\begin{abstract}
\medskip
The exploration of the phase diagram of Quantum ChromoDynamics (QCD) is carried out by studying ultrarelativistic heavy-ion collisions. The energy range covered by the CERN SPS ($\sqrt{s_{\rm \scriptscriptstyle{NN}}} \sim 6\text{--}17$~GeV) is ideal for the investigation of the region of the phase diagram corresponding to finite baryochemical potential ($\mu_{\rm B}$), and has been little explored up to now. We propose in this document a new experiment, NA60+/DiCE (Dilepton and Charm Experiment), that will address several observables which are fundamental for the understanding of the phase transition from hadronic matter towards a Quark--Gluon Plasma (QGP) at finite $\mu_B$. In particular, we propose to study, in Pb--Pb collisions, as a function of the collision energy, the production of thermal dimuons from the created system, from which one can obtain a caloric curve of the QCD phase diagram that may be sensitive to the order of the phase transition. In addition, the measurement of a $\rho\text{--}{\rm a}_1$ mixing contribution will provide conclusive insights into the restoration of the chiral symmetry of QCD.
Studies of open charm and charmonium production will also be carried out, addressing the measurement of transport properties of the QGP and the investigation of the onset of the deconfinement transition. Reference measurements with proton-nucleus collisions are an essential part of this program.
The experimental set-up couples a vertex telescope based on state-of-the-art monolithic active pixel sensors (MAPS) to a muon spectrometer with tracking detectors (MWPC). Two existing CERN dipole magnets, MEP48 and MNP33, the first being stored and the second currently in use by NA62, will be used for the vertex and muon spectrometers, respectively. The continuing availability of Pb ion beams in the CERN SPS is a crucial requirement for the experimental program. 
After the submission of a LoI , the experiment proposal is currently in preparation and is due by mid 2025. The start of the data taking is foreseen by 2029/2030, and should last about 7 years.

\end{abstract}
\vfill
{\centering\large\today\par}
\end{titlepage}

\setcounter{page}{1}
\section{Scientific context and objectives}
\label{sec:ScientificContext}
\bigskip

In the early universe, at about 10 $\mu$s after the Big Bang, 
a hot plasma of quarks and gluons (QGP) converted into massive hadrons. This transition has by now been well established by numerical simulations of a lattice-discretized QCD partition function, as a cross-over transition at a pseudo-critical temperature of $T_{\rm pc}\simeq$~155\,MeV~\cite{HotQCD:2018pds}. 

High-energy heavy-ion collisions at RHIC and the LHC have enabled detailed investigations of hot QCD matter. While the high-energy frontier of heavy-ion collisions probes the QCD medium at high temperatures and small baryo-chemical potentials, $\mu_B\simeq 0$, it is of high interest to investigate the regime of large baryon densities. Highly compressed nuclear matter at relatively low temperatures still exists in the universe today inside neutron stars and their mergers. Rather little is known about QCD matter at high $\mu_B$, but theoretical calculations suggest a potentially rich phase structure including the emergence of a first-order transition along with a second-order critical endpoint~\cite{Fukushima:2010bq}. By lowering the collision energies, heavy-ion experiments provide unique opportunities for systematic studies of a substantial part of the QCD phase diagram at high $\mu_B$ 
, thereby also promising to unravel connections between astrophysical systems and the early universe.  


At the CERN SPS, pioneering experiments with Pb-ion beams took place from 1994 to 2003, essentially investigating the higher-end of the energy range accessible to this accelerator ($\sqrt{s_{\rm NN}}\sim 17$ GeV)~\cite{Bass:1998vz,PRESSCUT-2000-210}. However, the full energy interval that can be explored at the SPS extends down to $\sqrt{s_{\rm NN}}\sim 5-6$ GeV, and was little explored, in particular for observables characterized by a relatively low production cross section. Experiments exploiting the full SPS coverage can study the region from $\mu_{\rm B}\sim 220$ up to about 500 MeV, 
likely approaching (from above) the region of the QCD phase diagram where the presence of the critical point is foreseen.

This collision energy region was recently explored by the STAR experiment at RHIC in the frame of the Beam Energy Scan program~\cite{STAR:2017sal}. The issues inherent to the operation of a collider machine much below its nominal working energy have limited the integrated luminosity that could be collected, so that mainly low-mass hadron observables have been studied. Nevertheless, measurements of net-proton number fluctuations in Au+Au collisions in the interval $7.7<\sqrt{s_{\rm NN}}<27$ GeV show intriguing deviations from model calculations that do not include the presence of a critical point~\cite{STAR:2020tga}.

In this context, the SPS, thanks to its possibility of accelerating high-intensity ion beams (from $\sim 20$ to 150 AGeV), provides an ideal environment for an accurate characterization of a high-$\mu_{\rm B}$ QGP. 
A class of observables that greatly benefits from a large integrated luminosity is the production of electromagnetic probes of the QGP. 
In particular, dileptons are a unique tool to determine the temperatures and lifetime of the strongly interacting medium created in the collisions. Those are key quantities to investigate a possible  anomalous behavior related to the onset of a first-order transition. In addition, dileptons provide direct information about hadron spectral functions, in particular the $\rho$-meson and its mixing with the chiral partner $a_{\rm 1}$, sensitive to chiral symmetry restoration which occurs in vicinity of the transition from hadrons to a QGP. Also for these observables, extensive studies were carried out only at top SPS energies, by the CERES~\cite{CERES:2006wcq} and NA60 experiments~\cite{Arnaldi:2006jq,Arnaldi:2008er,Arnaldi:2007ru}. In particular the latter experiment has carried out the first measurement of the $\rho$-meson spectral function in high-energy nuclear collisions, and provided evidence for thermal dilepton production, yielding the first direct measurement of a temperature of the system exceeding $T_{\rm pc}$. An extension of studies toward lower energies  presents a strong physics interest, due to expected sensitivity of the observables to $\mu_{\rm B}$ and to the approach toward the QCD critical point.

Another observable which profits of large integrated luminosity is heavy-quark production. It represents an excellent probe of the strongly interacting matter created in nuclear collisions and was never investigated below top SPS energy. The study of open charm production at collider energies has shown that charm quarks thermalize in the QGP and allowed an evaluation of the heavy-flavour diffusion coefficient~\cite{ALICE:2021rxa}. Furthermore, the hadronization process of charm quarks once the system goes below $T_{\rm pc}$ has shown unexpected features, with an enhancement of baryon/meson ratios in both pp and Pb--Pb collisions with respect to $e^+e^-$ data~\cite{ALICE:2020wfu}. An extension to lower energies is particularly interesting. The initial temperature of system will be lower and more time will be spent in the hadronic phase, allowing stringent tests of theory predictions for the diffusion coefficient in this region~\cite{Scardina:2017ipo}. The degree of thermalization of charm quarks may also be accurately measured, through studies of the elliptic flow of charmed hadrons. Finally, the hadronization of a baryon-rich QGP may lead to even larger enhancement of baryon/meson ratios with respect to LHC energies. 

Complementary to open charm, the investigation of charmonium states, and in particular the J/$\psi$, has always been of paramount importance in QGP studies. Anomalous suppression effects, likely due to color screening in the QGP, were first discovered at top SPS energies in the `90s (NA50/NA60 experiments)~\cite{Alessandro:2004ap,Arnaldi:2007zz} and subsequently investigated at RHIC/LHC~\cite{Adare:2011yf,Adam:2016rdg}. Moving to lower energies, suppression effects should finally disappear. It would be particularly interesting to correlate them with a precise measurement of the temperature of the fireball, and perform comparisons with the temperature dependence of the modifications of the inter-quark potential, which can be computed starting from lattice QCD results~\cite{Rothkopf:2019ipj}. 

This physics framework presents  excellent opportunities for a new experiment at the CERN SPS, utilizing high-intensity heavy-ion beams to investigate both hard and electromagnetic probes of the QGP through a systematic energy scan.
The observables detailed above can be studied with a set-up that includes a muon spectrometer, for the measurement of low-mass hadronic resonances ($\rho$, $\omega$, $\phi$), of charmonium states (J/$\psi$ and possibly $\psi{\rm(2S)}$ and $\chi_{\rm c}$), of the continuum related to thermal production from the QGP and the hadronic phase, and of its modifications connected with chiral symmetry restoration. The set-up must also include a high-resolution vertex spectrometer, to improve the accuracy of muon tracking, and to allow  secondary vertices from the decay of open charm mesons and baryons to be detected.

The project was established in 2016 within the Physics Beyond Colliders (PBC) initiative, adopting the "NA60+" designation to emphasize its continuity with the NA60 experiment. The latter collected data from In--In collisions in 2003 using a conceptually similar detector configuration.
An expression of interest~\cite{Dahms:2673280} was sent in 2019 to the SPSC, and a Letter of Intent~\cite{NA60:2022sze}  was presented to that committee in February 2023. Following the recognition of the fundamental interest of the measurement proposed, an encouragement toward the preparation of a proposal was issued~\cite{CERN-SPSC-2023-011}. The latter document is currently in preparation and its submission as NA60+/DiCE (Dilepton and Charm Experiment) is expected by mid 2025. 

\section{Methodology: experimental set-up and beam studies}
\label{sec:methodology}
\begin{figure}[ht]
\begin{center}
\includegraphics[width=1.0\linewidth]{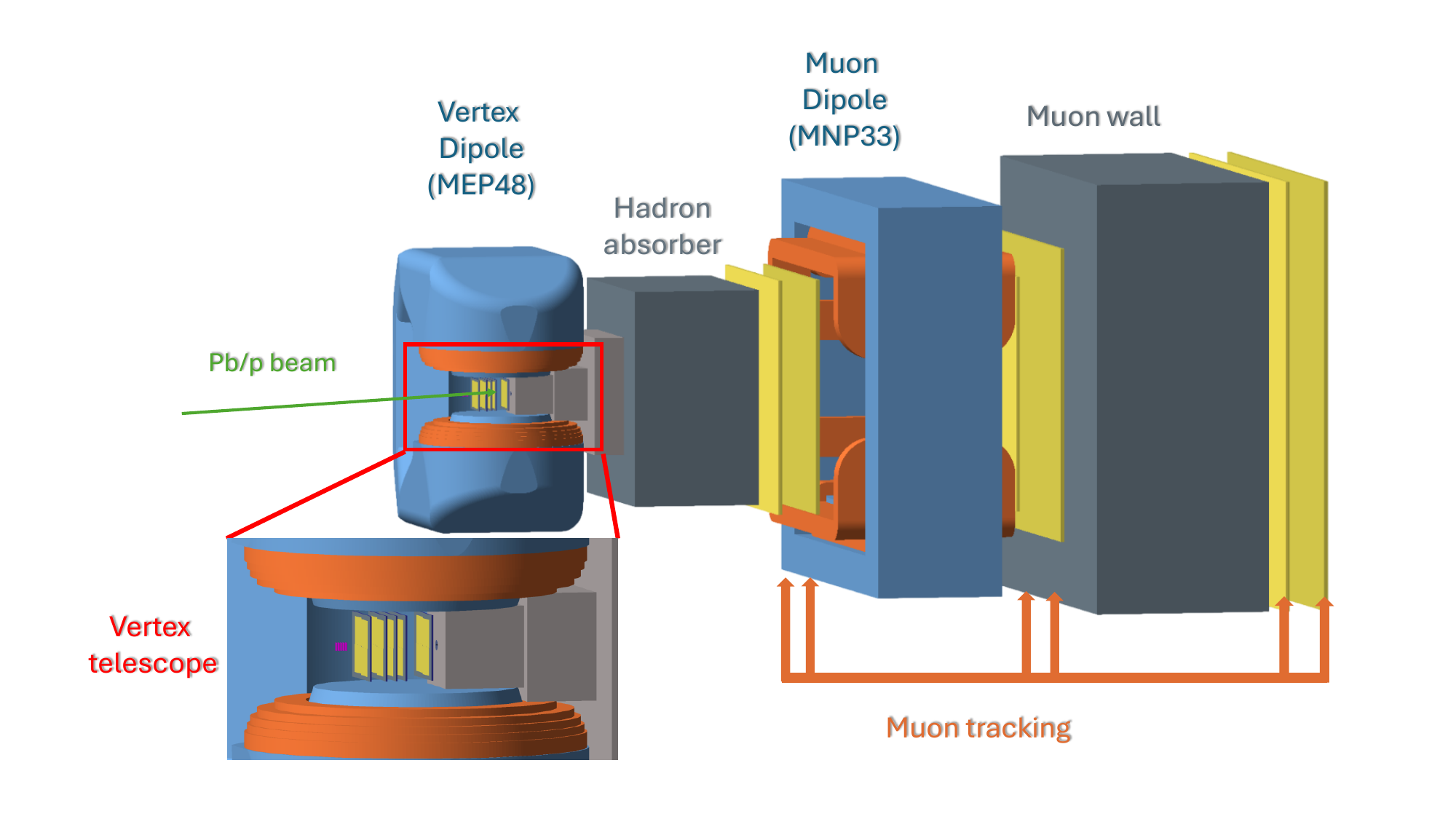}
\caption{GEANT4 rendering of the NA60+/DiCE experimental apparatus, for the low-energy configuration. }
\label{fig:setup}
\end{center}
\end{figure}

\bigskip
The proposed experimental set-up is sketched in Fig.~\ref{fig:setup}. A Pb-ion (proton) beam hits a segmented target, made of Pb (various nuclear species), with an interaction probability of $\sim15$\%. Particles generated in the interaction are tracked in a vertex spectrometer, made by Monolithic Active Pixel Sensors (MAPS), organized in five identical stations. 
Details are given in Sec.~\ref{sec:vertexspectrometer} below. The vertex spectrometer is positioned inside the 40 cm wide gap of the MEP48 magnet, that produces a $\sim 1.5$ T dipole field. The magnet is currently stored at CERN and will need a refurbishment in order to repair a short circuit in one of the coils.
The absorber begins at 45 cm from the target system, ensuring effective rejection of background muons from pion and kaon decays.
The total active area of the telescope is approximately 0.5 m$^2$.

\bigskip

Downstream of the vertex spectrometer region, a thick hadron absorber, made of BeO and graphite, selects muons out of the large multiplicity of charged particles (${\rm d}N/{\rm d}\eta >400$ for central Pb--Pb collisions at top SPS energy). The core of the hadron absorber is made of tungsten 
and dumps the uninteracting Pb or proton beam particles. The hadron absorber is followed by a muon spectrometer, made by six MWPC tracking stations, of increasing transverse dimensions. The magnetic field for the momentum measurement is generated by the MNP33 dipole. Its strength is $\sim 0.4$ T at the center of the $240\times 240$ cm$^2$ gap, but it extends along the beam direction well beyond the 130 cm length of the iron yoke, decreasing to 10\% of its maximum value at $\sim\pm2$ m in the z-direction. Details on the muon spectrometer are given in Sec.~\ref{sec:muonspectrometer} below. The two last tracking stations are preceded by a second absorber, made of graphite and 180 cm thick, that stops punch-through hadrons from the main absorber. A low- and a high-energy configuration of the muon spectrometer is foreseen, the latter featuring a thicker hadron absorber and a downstream shift of the detectors and magnet, to keep a constant rapidity acceptance. 
The original design in the LoI included the construction of a new toroidal magnet, as the MNP33 dipole was not available at the time. The subsequent availability of the MNP33 dipole provides both economic and technical advantages, particularly for low-momentum muons, since its open geometry (without axial coils near the beam axis) offers improved acceptance compared to a toroidal configuration.

\bigskip

\subsection{Vertex spectrometer}
\label{sec:vertexspectrometer}
\vskip 0.3cm
The primary function of the vertex spectrometer is to measure the kinematic parameters of muons before they reach the hadron absorber. Additionally, it detects and reconstructs hadronic tracks, enabling further studies such as open charm and strangeness production.

The requirements for the silicon sensor are: minimization of dead regions over the surface of a station; spatial resolution of 5 $\mu$m or better; material budget less than 0.1\% $X_0$ per silicon station; radiation hardness of $\sim 10^{14}$ 1 MeV n$_{\rm eq}$/cm$^2$; and sufficiently fast readout to cope with an interaction rate of 150 kHz in Pb-Pb collisions.

These requirements are optimally fulfilled by the new large-area monolithic active pixel sensor (MAPS) MOSAIX, which is currently being designed for the ALICE ITS3 and is based on the 65 nm technology developed by Tower Partners Semiconductor (TPSCo)~\cite{The:2890181}. A process called stitching is used to manufacture sensors with diagonals that can be up to the diameter of a 300 mm wafer, overcoming the size limitation of the design reticle, which typically measures approximately 3 x 2 cm$^2$. A preliminary layout of the photomask is structured into subunits. The reticle includes three sub-designs: the RSU (Readout Sensor Unit), which is a pixel matrix, and the left and right end caps (LEC, REC).

The RSU measures 19.564 $\times$ 21.666 mm$^2$ and consists of two mirrored half-sensor units (top and bottom), each segmented into tiles. Each tile contains four readout regions and functions as an independent sensor with separate configuration and a serial readout link operating at 160 Mbps. The digital pixel matrix in the readout regions is 460 $\times$ 162 pixels, with a pixel pitch of 20.8 $\times$ 22.8 $\mu$m$^2$.
The pixel architecture features: a continuously active front-end; in-pixel discrimination; a global shutter for simultaneous recording of all hit pixels; zero-suppressed matrix readout; and continuous readout mode. The LEC is a physical interface between tiles and the off-chip domain. It has two main functions: to provide all control signals to tiles, including clock and reset,  and to collect data coming from tiles and send it off-chip. The REC provides additional power pads needed to compensate for voltage drops across the sensor.

\begin{figure}[!ht]
  \centering
\includegraphics[width=0.35
\textwidth]{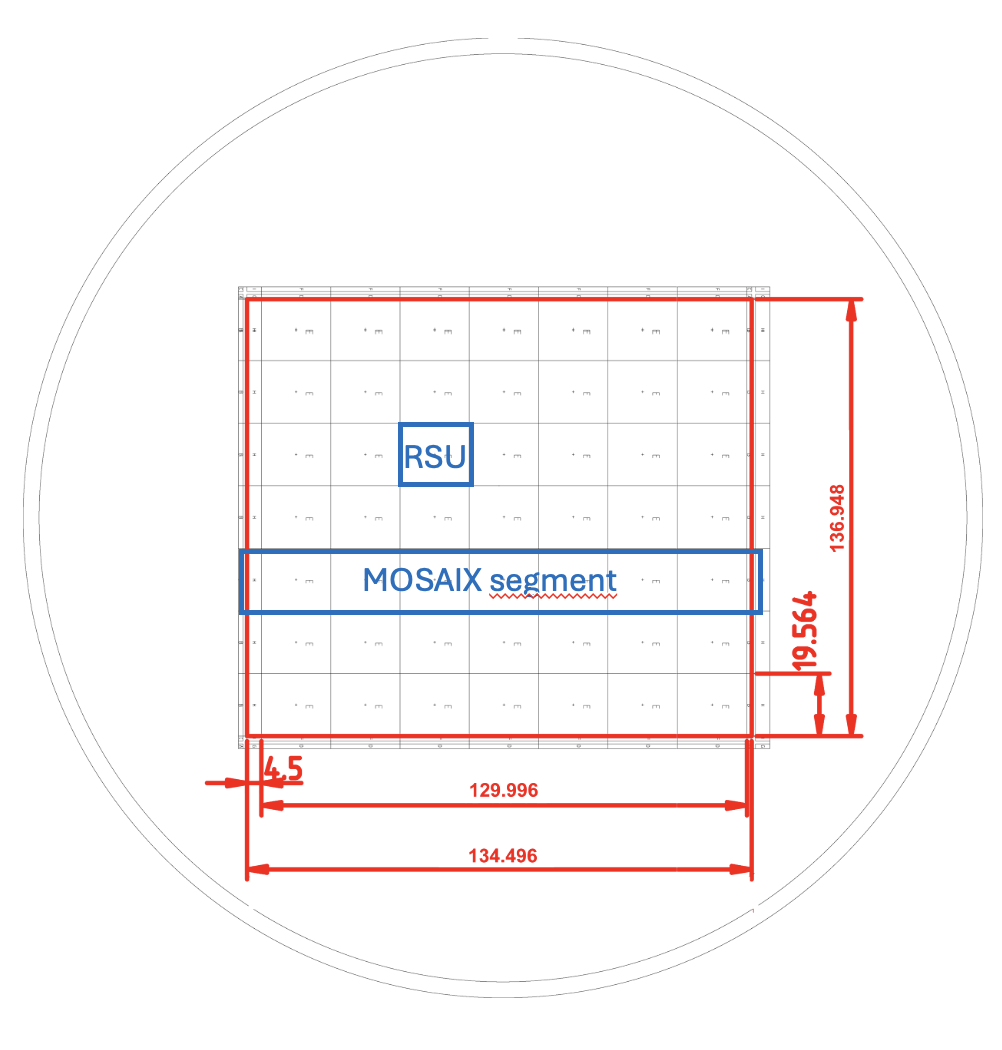}
\includegraphics[width=0.62\textwidth]{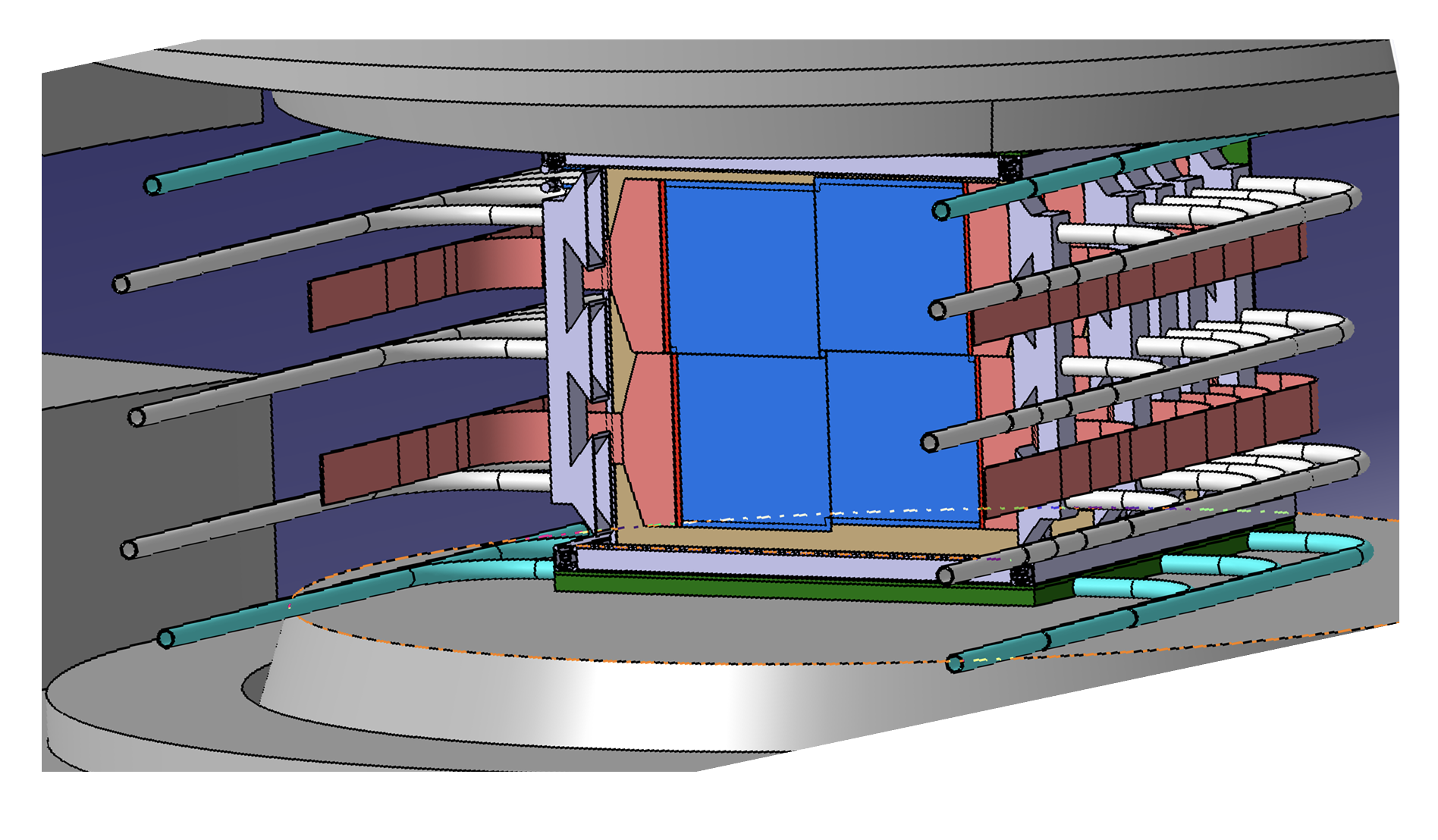}
 \caption{Stitching plan of the NA60+/DiCE sensor (left); the pixel stations with 4 sensors integrated inside MEP48 (right).}
  \label{fig:Mosaix}
\end{figure}

The RSUs are repeatedly lithographed six times onto the wafer in adjacent locations according to a predefined pattern with accurate translations and alignment, as shown in Fig.~\ref{fig:Mosaix}(left panel).
This defines a so-called MOSAIX segment.
The subsequent deposition of an additional stack of metal bridges, or stitches, connects otherwise unconnected reticles on the wafer during the production stage. Along with the supply lines, each tile's data output has a direct connection to the LEC through the stitched backbone, where all I/Os from a full row of RSUs are fed into eight differential data output serializers operating at 10.24 Gbps.

The stitching plan to fabricate a 13.6 $\times$ 13.6 cm$^2$ sensor on a 300 mm silicon wafer is also shown in Fig.~\ref{fig:Mosaix}(left panel). The MOSAIX segment of 6 RSUs stitched horizontally is repeated 7 times vertically. This is the basic sensor for NA60+/DiCE. With a thickness of 50 microns, it features a material budget of less than 0.1\% X$_0$. The pixel pitch will provide a spatial resolution of approximately 5 microns. The sensor's maximum particle rate is 6 MHz/cm$^2$, which has been shown to be adequate based on detailed simulations of Pb-Pb interactions using Fluka.

The pixel stations are formed by 4 sensors, as shown in Fig.~\ref{fig:Mosaix} (right panel), integrated inside the MEP48 magnet.
The sensors are positioned to leave a central square hole of 6 mm in size for the beam passage. This configuration highlights the advantage of using large-area sensors, as they significantly reduce dead zones that would inevitably be present if small-area sensors were used. 

The power consumption of the pixel matrix is 40 mW/cm$^2$. Furthermore, the LEC, hosting the serial drivers, has a power consumption of 791 mW. A cooling system is required to keep the sensor's temperature around 25 degrees. 
The silicon sensors are glued onto a 400-micron-thick carbon fiber plane, which serves for efficient heat extraction and exchange with the surrounding air. The silicon stations are hosted inside an  aluminum box with tubes to establish an air flux (shown in gray in the right panel of Fig.~\ref{fig:Mosaix}). Air at 18 degrees Celsius is flushed at 2 m/s tangent to the sensor surface to remove the heat. The box will also provide insulation from the warm magnet expansions. The LEC power consumption requires an additional water cooling. For this purpose, the carbon fiber plate is glued onto an external aluminum frame in which a water beam pipe runs along the sensor peripheries (shown in green in the right panel of Fig.~\ref{fig:Mosaix}). 

\bigskip
\subsection{Muon spectrometer}
\label{sec:muonspectrometer}
\vskip 0.3cm
The primary function of the muon spectrometer (MS) is to measure the kinematic parameters of tracks penetrating the absorber(s). Matching them between the MS and \vrtx allows for selecting muon candidates. 

The current MS configuration is shown in Fig.~\ref{fig:setup}. It consists of the dipole analyzing magnet MNP33~\cite{Griesmayer:1995wd,Fry:2016sjj}, sandwiched between two pairs of tracking stations. Each station measures particle hit position. The difference between the angles of particles before and after the magnet determines the track kinematics, which can be matched to the tracks measured in the \vrtx, identifying muons. Two additional tracking stations located behind the additional absorber, indicated as the Muon wall in the Figure, allow for control of the purity of the muon selection and, if needed, to trigger on the presence of a muon in an event. 

The area of the MS stations corresponds to the opening of the MNP33 magnet,  $2.4\times2.4$~m$^{2}$. The required resolution of the MS detectors determined from the simulation is approximately 200~$\mu$m in the direction perpendicular to the MNP33 analyzing field. The two stations behind the Muon wall have larger coverage, approximately $4.5\times4.5$~m$^{2}$, and need coarser resolution. Since two setup configurations, low- and high-energy, are considered for the data taking, the whole MS spectrometer, including the MNP33 magnet, the Muon Wall, and MS stations, must be movable by several meters along the beam direction.

The accuracy of matching the track in the MS spectrometer to the \vrtx is primarily determined by the energy loss fluctuations of the muon passing through the Hadron absorber, shown in Fig.~\ref{fig:setup}. Improving the MS resolution beyond 200~$\mu$m only marginally improves the muon matching but requires more electronic channels in the MS. Other factors, like the detector material thickness, play a lesser role. A tungsten plug inside the Hadron absorber, which plays the role of a beam dump, greatly reduces the fluence of particles originating from the beam remnants that can reach the MS system. The flux of charged particles on the first MS station peaks around $2\times10^{3}$~cm$^{-2}$s$^{-1}$.

\begin{figure}[!ht]
  \centering
\includegraphics[width=0.3\textwidth]{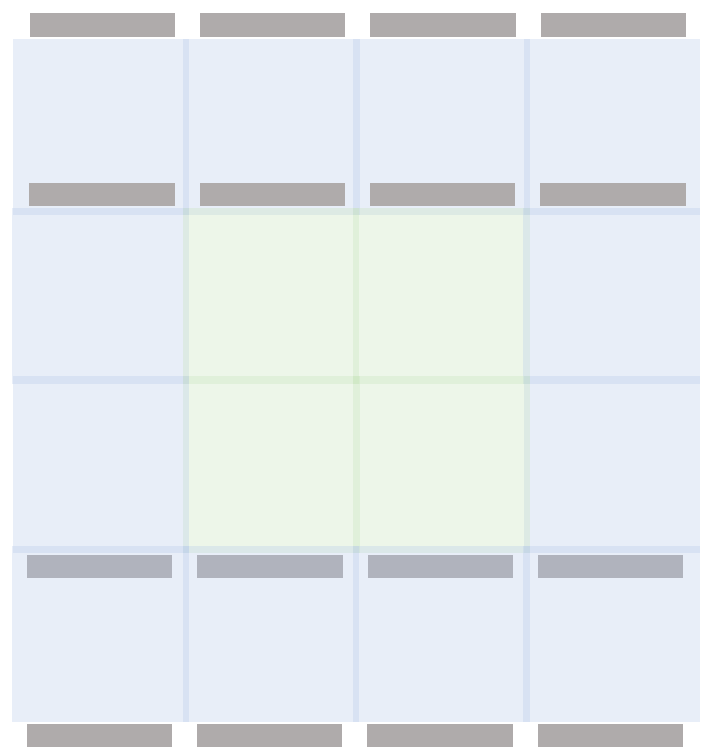} 
\hspace{5mm}
\includegraphics[width=0.3\textwidth]{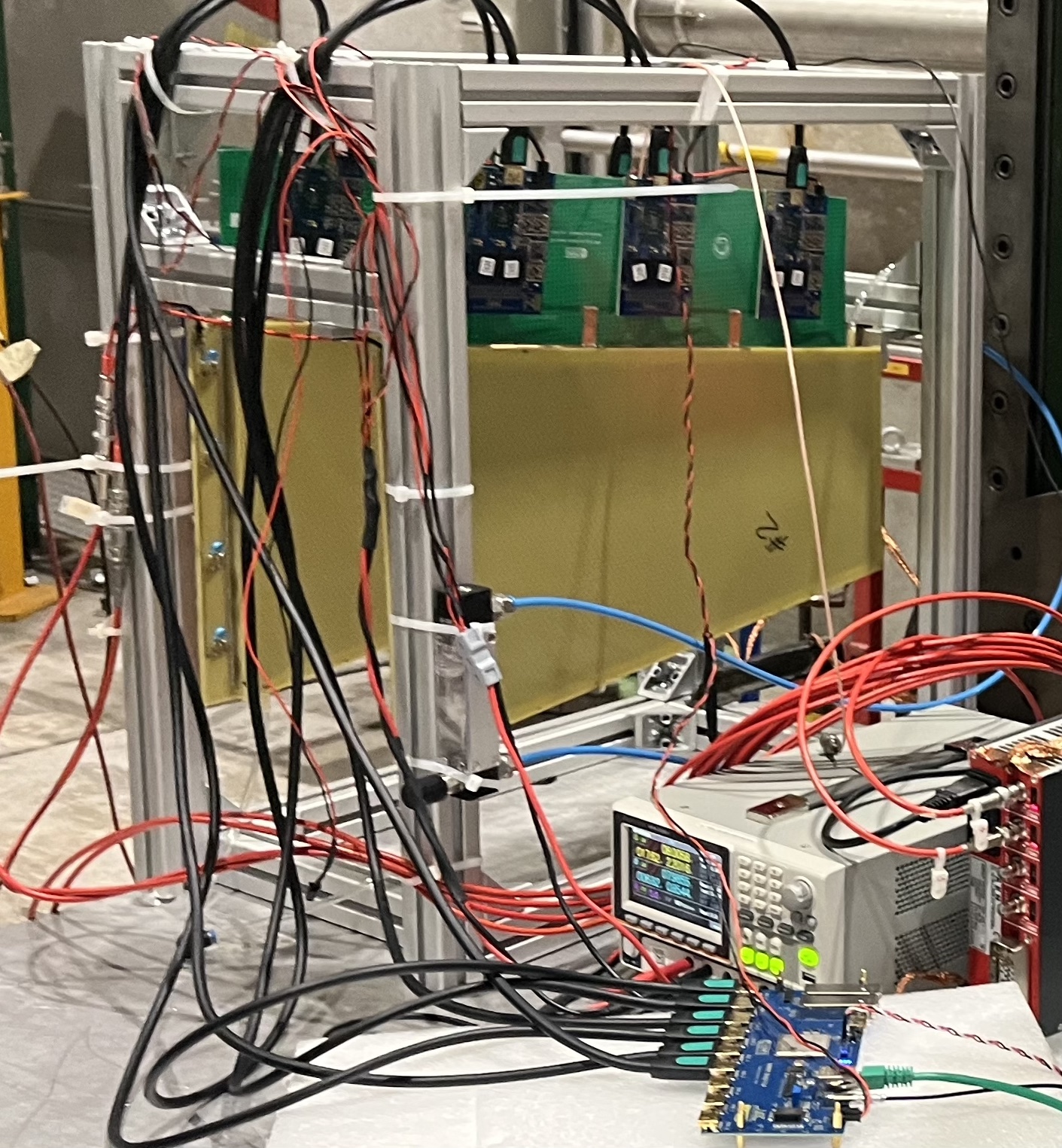}
 \caption{Left: Schematic layout of the MS readout modules in a MS tracking station. Two colors show sensitive areas that overlap at the edges where the color is darker. Different colors correspond to different readout channel densities, higher at the center. Grey strips indicate the position of the readout electronics. Right: Readout module prototypes during testing at CERN.}
  \label{fig:mwpc_modules}
\end{figure}

Considering these aspects, the technology choice for the MS tracking stations is made in favor of the multi-wire proportional chambers with strip readout. This technology allows building a highly reliable, large area, nearly 100\% efficient detectors at low cost, which fully satisfy the physics requirements of MS. The MS tracking stations shown in Fig.~\ref{fig:setup} will be built from smaller modules of only two or three different types, varying mainly in the number of readout channels. 
A conceptual layout of the MWPC modules in an MS station is shown in Figure~\ref{fig:mwpc_modules} (left panel).

Two prototypes of the readout modules were constructed and tested at CERN. The first used a two-sided readout strip system, and the second (see right panel of Fig.~\ref{fig:mwpc_modules}) used the single-sided strip system, based on the idea suggested in~\cite{Bressan:1998jj}. The single-sided system has advantages in solving ambiguities in hit association when the detector occupancy increases. The first prototype was tested with standard VMM3A readout electronics used by the Scalable Readout System data acquisition used in the first test. The second prototype used the first version of the VMM3A-based electronics planned for the experiment. The prototypes were about $1/3-1/2$ of the real size and were built in a trapezoidal shape. 
They demonstrated position resolution better than required for the MS stations. An efficiency of MWPC close to 100\% can be achieved with these chambers, but this needs to be tested with the final electronics currently in development. Since the MWPC technology is well established in many previous experiments, the prototype has had the optimization of the strip system as a goal for handling multi-hit events, testing some production steps, and identifying the vendors. The full-scale, nearly final prototype will be tested at CERN in 2025.

\bigskip
\subsection{Data acquisition and computing}
\vskip 0.3cm


The Data Acquisition (DAQ) system integrates the vertex spectrometer and the muon spectrometer. It comprises custom electronics for each subsystem and a common integration scheme. The DAQ will be designed to take data continuously at a rate of 150 kHz and process it online. The two subsystems are integrated into the global scheme using back-end readout cards, which are already available for other experiments at the Large Hadron Collider (LHC), e.g. the Common Readout Unit (CRU) \cite{CRU:alicels2} or the FELIX card \cite{FELIX:2023chep} hosted in dedicated processing servers. 
The back-end card receives the detector data through multiple optical links,  performs initial data formatting and, if needed, processing before sending it to the online processing farm. 


The data acquisition of the vertex spectrometer is done at the level of sensor segments as described in Sec.~\ref{sec:vertexspectrometer}. There are a total of 140 segments in the vertex spectrometer. Each  segment ships out data through multiple 10.24\,Gbps data links. Flexible printed circuits connect the segments to a Segment Readout Unit (SRU). The SRU hosts the Low Power Gigabit Transciever (lpGBT) ASIC \cite{LPGBT:2022man} and an optical link module called the VTRX+ \cite{VTRX:2017man}, implementing a high speed bidirectional optical communication interface between the detector front-end and the back-end electronics. The lpGBT and VTRX+ are developed at CERN for future high-luminosity experiments. The SRU also hosts voltage regulators to supply voltages to the sensor. The optical fibers from the VTRX+ connectors are connected to the common back-end CRU/FELIX card. The scheme satisfies the readout rate requirement of the NA60+/DiCE silicon tracker of $\sim 60$\,Gbps.

The data acquisition of the muon spectrometer is done at the level of MWPC modules. There are a total of 240 modules. Front-end cards (FEC) with two VMM3 \cite{vmm3:2018atlas} chips are connected directly to the detector modules. The VMM3 chips process the signals and send data to Data Collection Boards (DCB) via HDMI cables. Data from the DCB e-links are collected by Module Readout Units (MRU) which host the lpGBT ASICs and the VTRX+ modules. The MRUs forward the data to the backend CRU/FELIX card by optical links. Modules in the central region of the first muon tracking station receive a hit rate up to $6\times10^{5}$\,s$^{-1}$, whereas modules in the last  station about $10^{4}$\,s$^{-1}$. Modules have strip readout, and signals are digitized using a VMM3 chip, with 38 bits of data per channel per hit\,\cite{Wang_2020}, producing a data flow between 10 and 230\,Mbps, resulting from 7000 to 900\,Mbps from the muon tracking stations. These rates are small compared to the scheme's throughput. The total data flow generated by the muon system in a non triggered regime is less than 15\,Gbps. 


The readout data rate as well as the data storage and processing requirements of NA60+/DiCE are fully dominated by the contribution from the vertex telescope.
Since the code for the simulation of the readout of MOSAIX sensors is still under development, we pessimistically assume 58 bits per particle hit at the raw data readout level,
as measured in ALICE Pb-Pb data of the Muon Forward Tracker (MFT)~\cite{MFT:alice2} with ALPIDE sensors (accounts for the hardware partial data compression and overheads of the data formatting).

According to Fluka simulations for a $40$ AGeV Pb beam impinging on the $15\%$ interaction probability we should expect 
in the vertex telescope on average $\sim 2.1\cdot 10^3$ hits in the case of a hadronic interaction taking place in one of the targets and $\sim 8.3\cdot 10^2$ hits otherwise (due to the $\delta$-rays and conversions from electromagnetic interactions). This would lead in average to $\sim 1.03\cdot 10^3$ hits per incoming ion if we would use a trigger-less continuous readout. With the expected intensity of $\sim 10^7$ ions per spill this means $\sim 600$ Gbit/spill.

Collected raw data will be stored on a temporary disk buffer and immediately pre-processed. This initial processing will allow its compression to the level of $\sim 21 $ bit/hit using the loss-less compression method already validated in ALICE for ITS and MFT data: raw data are decoded, fired pixels clusterized and for each cluster the reference channel (containing the center of gravity of the cluster) together with cluster topology identifier is compressed using the highly optimized version of rANS algorithm implemented by ALICE \cite{lettrich:AliceRANS}.
With the planned  yearly statistics of $\sim 0.6\cdot 10^{12}$ Pb ions on target, this would amount to $\sim 4.5$ PB/year of collected raw data and   to $\sim 1.7$ PB/year of archived data.
The compressed data for each period will be held on a $\sim 2$ PB disk storage for one year until they are fully processed and also will be immediately copied to the tape storage. 

\bigskip
\subsection{Pb and proton beams}
\vskip 0.3cm

The study of the proposed observables requires high beam intensities. The results of the physics performance studies described in the LoI were based, at each collision energy, on 10$^{12}$ Pb ions incident on a 15\% interaction probability (segmented) Pb target. The corresponding integrated luminosity should be collected in the typical time period devoted to heavy-ion studies in one year, of about 4 weeks.  
Detailed studies were carried out in the frame of a PBC study group~\cite{Bernhard:2925964}, to assess a realistic scenario for the delivering of Pb beams to NA60+/DiCE in the PPE138 experimental hall on the H8 beam line. 
The results show that radiation protection consideration are the dominant factor, and limit the total number of Pb-ions on target to about $0.6\times10^{12}$. Although smaller than the number considered in the LoI, other modifications in the set-up, and in particular the use of the MNP33 dipole instead of the initially foreseen toroid, brought to an increase of acceptance that roughly compensates this decrease. 

Another required feature for the beam is the need for a sub-mm size, due to the fact that the geometry of the vertex spectrometer stations defines a square central hole of $6\times6$ mm$^2$. This requires a careful design of the beam optics. Extensive tests~\cite{Bernhard:2925964} have been carried out with a 150A GeV Pb beam in the PPE138 zone, using a set-up made by several planes of ALPIDE Si pixel detectors to perform an accurate measurement of the beam features. A transverse size of the beam spot of $\sim 0.2\times0.2$ mm$^2$ (1$\sigma$) for a beam intensity of 10$^7$ Pb ions per spill was measured, well within the size required by the experiment. When moving toward lower energy the emittance of the beam naturally increases by $1/\sqrt{p}$. Also, more sensitivity to the presence of residual material along the beam line is expected. 
A preliminary test conducted in 2024 using a 13.5A GeV Pb beam achieved a beam spot size of approximately $1\times 1$ mm$^2$, though further optimization appears achievable.

In addition to the Pb beam, NA60+/DiCE will need proton beams~\cite{Scomparin:2925719}. The study of p--A collisions, at the same energy per nucleon of the ion data, serves as a mandatory reference for the interpretation of Pb--Pb results. An integrated luminosity per nucleon-nucleon collision similar to that available for Pb--Pb is the standard choice. Extensive studies were carried out~\cite{Dyks:2923184}, to assess the possibility of using secondary beams coming from the fragmentation in the T4 target of a primary 400 GeV proton beam extracted from the SPS. However, due mainly to beam purity issues, this solution cannot be adopted.
The currently preferred option is to use a low-energy primary proton beam. However, hardware interlocks in the machine prevent the extraction of proton beams with momenta other than 400 GeV to the North Area during standard proton-running periods. Consequently, this solution is only viable during heavy-ion running periods, when the maximum extracted beam intensity is reduced to 10$^{11}$ charges per spill. The feasibility of this approach remains under investigation.
If this approach proves feasible, the experiment would collect data at a limited set of proton energies (tentatively three points between 40 and 150 GeV). The required reference results for all studied Pb beam energies would then be obtained through interpolation. This strategy would help mitigate scheduling issues arising from the need to deliver these low-energy beams to the entire North Area.
\section{Benchmark observables/measurements}
\bigskip

\subsection{Electromagnetic probes}
\bigskip
NA60+/DiCE proposes to perform the first measurement of a caloric curve for the phase transition between hadronic matter and the QGP. This measurement will determine the medium temperature as a function of collision energy using a precise dimuon thermometer, which is unaffected by the blue shift effect observed in momentum spectra. For thermal dilepton pairs with masses above 1.5 GeV/c$^2$, the yield is proportional to $dN/dM \propto M^{3/2} \exp(-M/T_{\text{slope}})$~\cite{Rapp:2014hha}, which depends only on mass and is thus by construction Lorentz-invariant, i.e., immune to any collective motion of the expanding source.
The parameter $T_{\text{slope}}$ in the shape of the mass spectrum is a space-time average of the time-dependent temperature $T$ during the fireball evolution. The choice of the intermediate-mass region (IMR), 1.5--2.5 GeV/c$^2$, implies $T \ll M$ and thus strongly enhances the sensitivity to the early high-temperature phases of the evolution.

The experimental program  proposes to perform an energy scan in the interval $\sqrt{s_{\text{NN}}} = 6-17$ GeV ($E_{\text{lab}} = 20-150$ AGeV), with particular focus on $\sqrt{s_{\text{NN}}} < 10$ GeV, which is believed to be essential to map out the phase transition regime at high $\mu_B$. This could lead to the discovery of a plateau in the caloric curve obtained with dilepton slopes $T_{\text{slope}}$. With an experimental precision on $T$ of a few MeV, as targeted by NA60+/DiCE, the experiment will have excellent capability to identify the transition region in this critical part of the QCD phase diagram.

\begin{figure}[h]
\begin{center}
\includegraphics[width=0.45\textwidth]{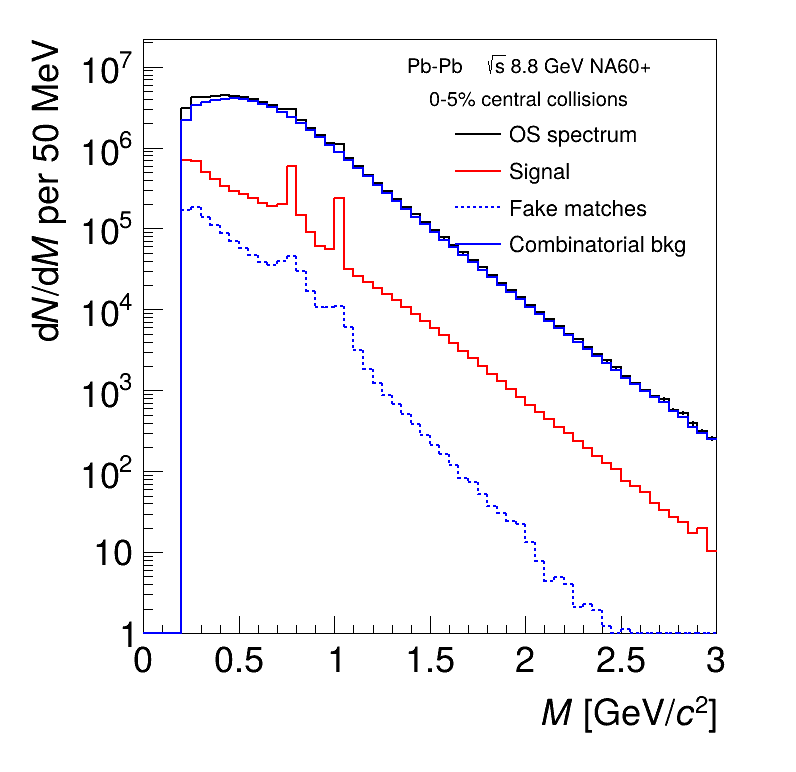}
\includegraphics[width=0.45\textwidth]{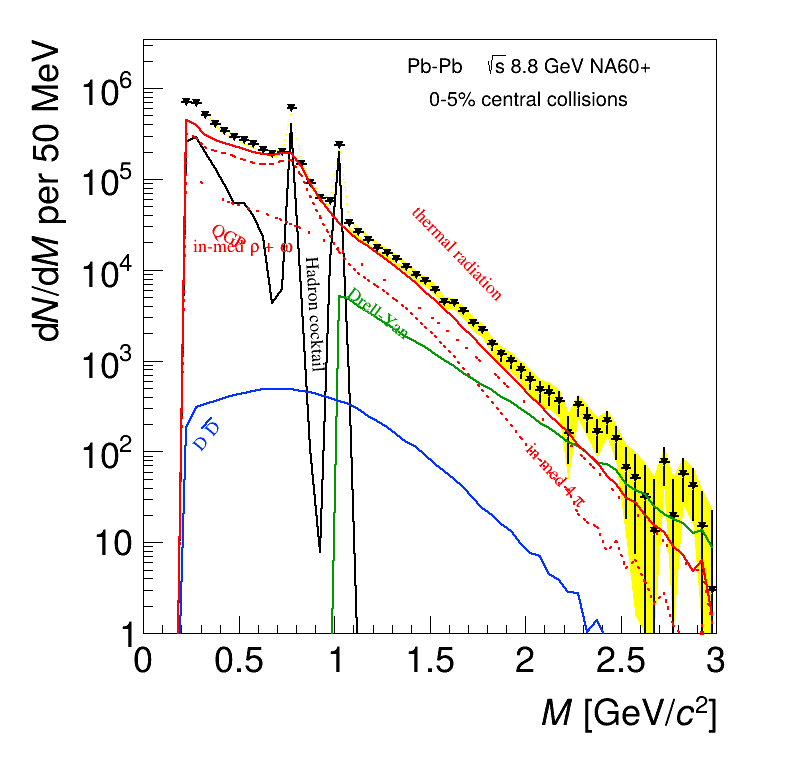}
\caption{Expected dimuon sample in the 5\% most central \PbPb collisions at $\sqrtsNN = 8.8\GeV$ before (left) and after (right) 
subtraction of combinatorial and fake match background.
Various contributions are shown (see text for details). }
\label{fig:DileptonMassSpectra}
\end{center}
\end{figure}

Here, we present detailed performance studies for the 5\% most central Pb-Pb collisions at $\sqrt{s_{\text{NN}}} = 8.8$ GeV. The differential spectra of thermal $\mu^+\mu^-$ pairs, $d^3 N/dM d\pt dy$, are based on the in-medium $\rho$, $\omega$, and 4-pion spectral functions, QGP radiation, and the expanding thermal fireball model of~\cite{Rapp:2014hha}. 
The results correspond to a data sample collected in one month of data taking,
 with a total integrated luminosity of $0.6 \times 10^{12}$ ions on target. 
Fig.~\ref{fig:DileptonMassSpectra} (left panel) shows the total reconstructed mass spectrum (black).
The combinatorial background (continuous blue line) is estimated using Fluka simulations, which take into account hadronic interactions in the absorber and the muon wall. The average signal-to-background ratio at $M = 0.6$ GeV/c$^2$, in a region completely dominated by the thermal contribution, is $\sim 1/20$.  The net signal after subtraction of the combinatorial background and fake matches is shown in red. The contribution of the fake matches is small compared to the combinatorial background, becoming completely negligible for $M > 1$ GeV/c$^2$.

Fig.~\ref{fig:DileptonMassSpectra} (right panel) shows the reconstructed signal mass spectra (black) for Pb-Pb collisions at $\sqrt{s_{\text{NN}}} = 8.8$ GeV after subtraction of the combinatorial background due to pion and kaon decays as well as fake matches. The 0.5\% systematic uncertainty from the subtraction of the combinatorial background is shown as a yellow band. The figure shows all the expected signal components. For $M < 1$ GeV/c$^2$, the thermal radiation yield is dominated by the in-medium $\rho$. The $\omega$ and $\phi$ peaks are well resolved, with a resolution better than 10 MeV/c$^2$ at the $\omega$ mass. The thermal spectrum is measurable up to 2.5--3 GeV/c$^2$. The open-charm yield becomes totally negligible at low $\sqrt{s_{\text{NN}}}$. The Drell--Yan yield will be measured in dedicated p-A runs.

\begin{figure}[h]
\begin{center}
\includegraphics[width=0.45\textwidth]{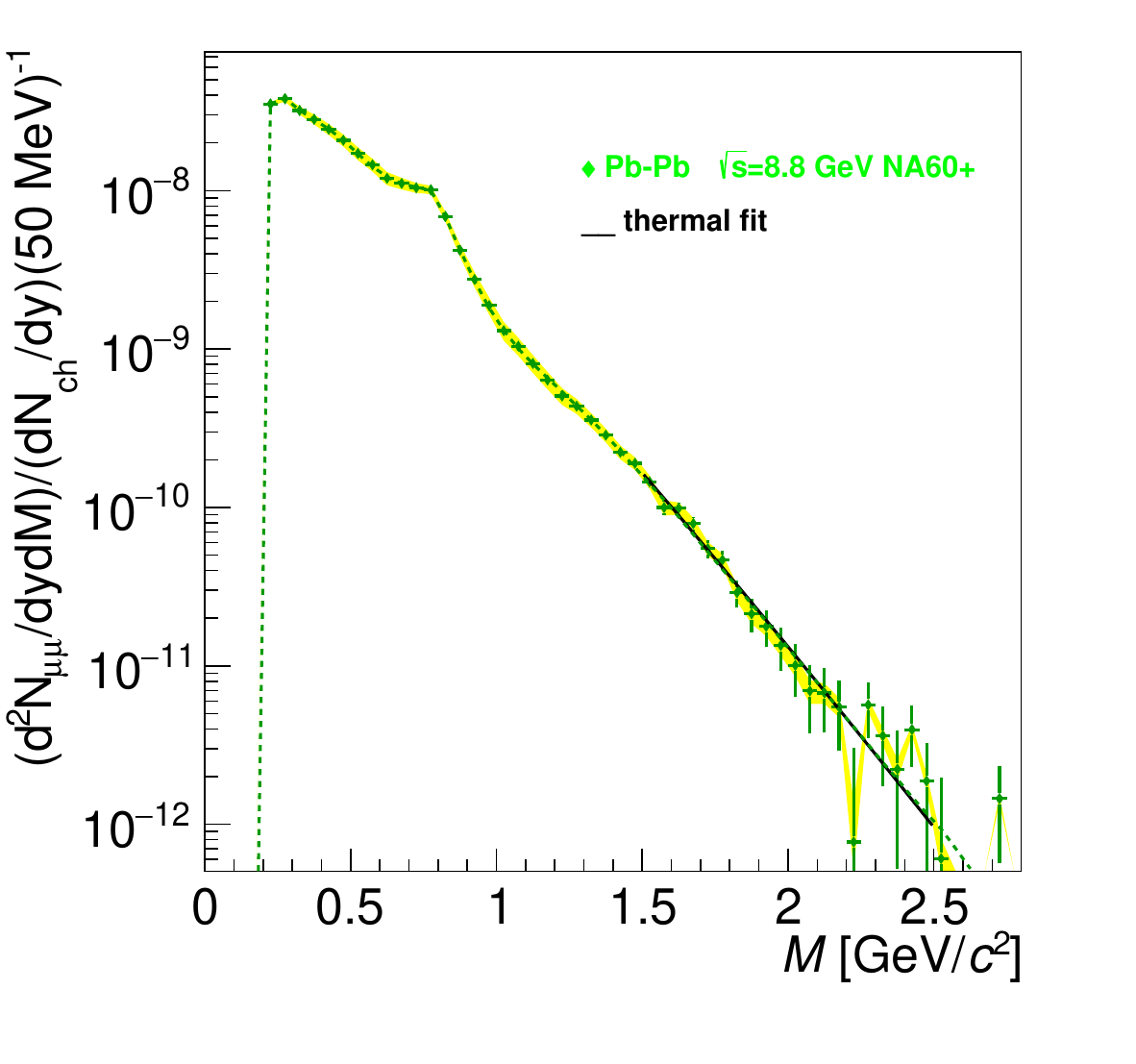}
\includegraphics[width=0.45\textwidth]{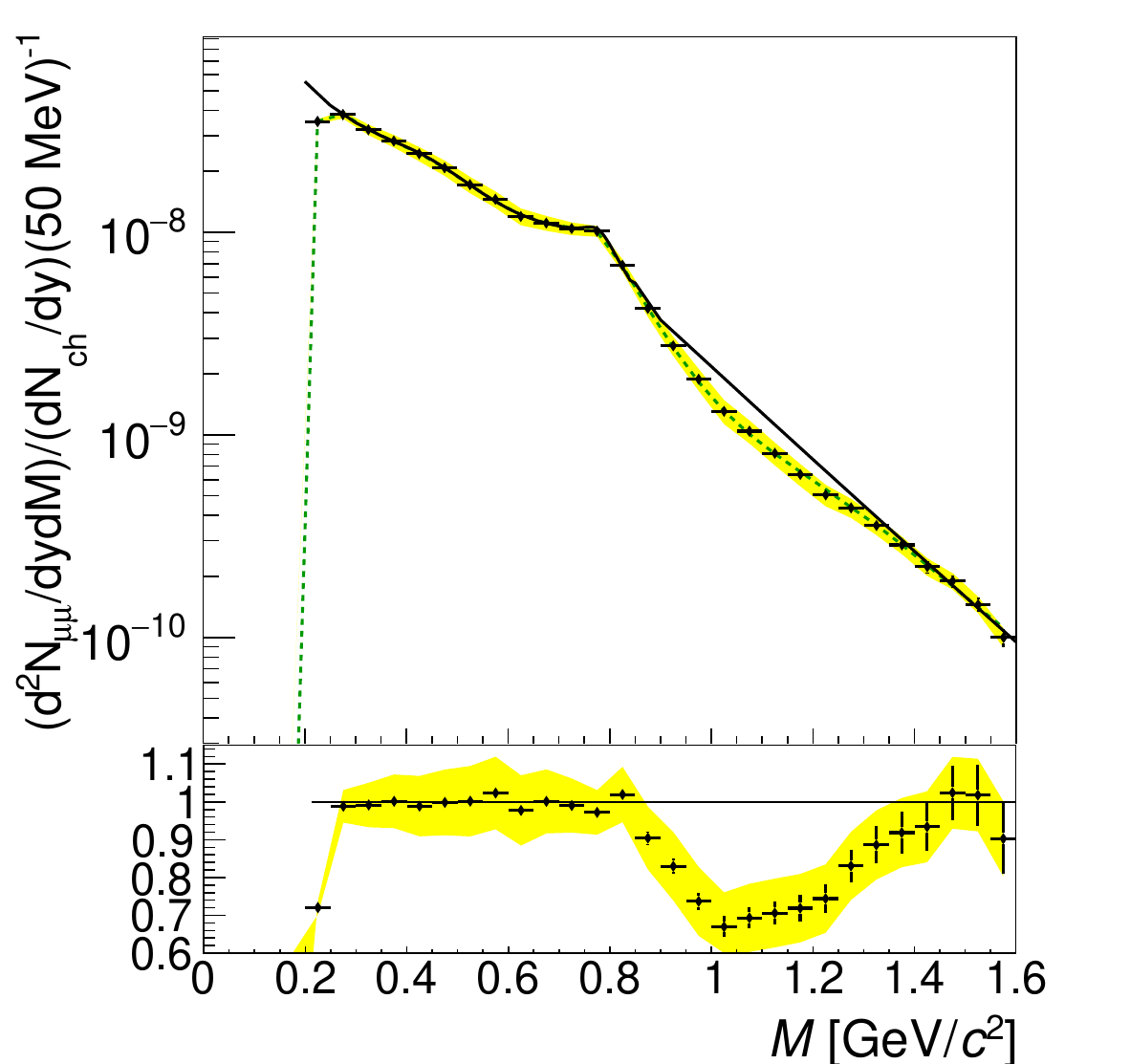}
\caption{
(Left) Acceptance corrected thermal spectrum at $\sqrtsNN = 8.8 \GeV$ 
Model comparisons and exponential fits as discussed in the text are shown.
(Right) 
Acceptance corrected thermal dimuon mass spectrum at $\sqrtsNN = 8.8\GeV$ in case of no chiral mixing compared to the theoretical expectation from full chiral mixing (black line)~\cite{Rapp:2014hha}.
}
\label{fig:fig1-thermal-performance}
\end{center}
\end{figure}

The thermal spectrum is obtained after (i) subtraction of the hadronic cocktail for $M < 1$ GeV/c$^2$, including $\eta$, $\omega$, and $\phi$ decays into $\mu^+\mu^-$ as well as the $\eta$ and $\omega$ Dalitz decays, and (ii) subtraction of Drell--Yan as well as open-charm muon pairs for $M > 1$ GeV/c$^2$. After acceptance correction, the spectrum is fit with $dN/dM \propto M^{3/2} \exp(-M/T_{\text{slope}})$ in the interval $M = 1.5-2.5$ GeV/c$^2$. The resulting spectrum is shown in Fig.~\ref{fig:fig1-thermal-performance} (left panel). The theoretical spectrum used as input is shown as a dashed line, while the exponential fit is shown as a black line.
The temperature extracted from the fit is \( T = 169 \pm 7 \pm 2 \) MeV. The combined statistical and systematic uncertainties result in a 4\% precision on \( T \), demonstrating the experiment's strong sensitivity to a potential flattening of the caloric curve in a region complementary to that which will be explored by the CBM experiment at FAIR/GSI.

The broadening of the $\rho$-meson spectral function observed at SPS~\cite{CERESNA45:1997tgc,CERESNA45:2002gnc,CERES:2006wcq,Arnaldi:2006jq,Arnaldi:2008er,Specht:2010xu} and later at RHIC energies~\cite{STAR:2013pwb,PHENIX:2015vek} is consistent with chiral symmetry restoration. However, an unambiguous observation of chiral symmetry restoration would require measuring not only the $\rho$-meson spectral function but also that of its chiral partner $a_{\rm 1}$. Unfortunately, the latter cannot be reconstructed exclusively in heavy-ion collisions.

The so-called $\rho - a_{1}$ chiral mixing mechanism provides indirect access to the properties of the $a_{\rm 1}$. In essence, the presence of pions in the surrounding medium (both real and virtual) leads to an "admixing" of the axial-vector channel into the vector channel (and vice versa), producing dileptons. In particular, $\pi + a_1$ annihilation processes are most prominent in the ``dip region'' of the vector spectral function in vacuum, for masses $M \simeq 0.9-1.4$\,GeV.
As the temperature increases, this dip is filled, and near the pseudo-critical temperature the vector spectral function essentially flattens out, signaling the approach to chiral symmetry restoration \cite{Hohler:2013eba}. Although this effect is relatively small, the change in this region is sensitive to the mixing of the chiral partners $\rho$ and $a_1$, and thus to chiral symmetry restoration. 

The mixing effect can be identified through a precise experimental study of the thermal dilepton spectrum in the invariant mass region between 0.9 and 1.4 GeV. The acceptance corrected mass spectrum,
based on the assumption of no chiral mixing, is compared to the expectation of full chiral mixing in  Fig.~\ref{fig:fig1-thermal-performance} (right panel).
As shown, the statistical and systematic uncertainties provide a very good sensitivity to an increase of the yield due to chiral mixing of ${\sim}20\text{--}30\%$.


\subsection{Open charm}
\bigskip

Open charm hadrons can be fully reconstructed with the NA60+/DiCE apparatus through their decays into two or three charged hadrons. Specifically, the following decays allow for the measurement of non-strange and strange D mesons as well as $\lambdac$ baryons: $\Dzero \to {\rm K}^-\pi^+$, $\Dplus\to {\rm K}^-\pi^+\pi^+$, ${\Ds \to \phi\pi^+\to {\rm K}^{+} {\rm K}^{-} \pi^{+}}$, $\lambdacplus \to {\rm p K^{-}}\pi^+$, and their charge conjugates. 
The decay products of the charm hadrons (pions, kaons, and protons) are detected by reconstructing their tracks in the vertex telescope.
D-meson and \lambdac candidates are formed by combining pairs or triplets of tracks with the correct charge signs. The large combinatorial background is reduced through geometrical selections based on the displaced decay-vertex topology, exploiting the fact that the mean proper decay lengths $c\tau$ of open charm hadrons range from 60 to 310~$\mu$m depending on the hadron species. 

Benchmark studies were performed for the measurement of the two-body decay of \Dzero mesons and the three body decays of \Ds mesons and \lambdacplus baryons in the 5\% most central \PbPb collisions at the top SPS beam energy of 150 $\GeV/\text{nucleon}$, corresponding to $\sqrtsNN = 16.8$ \GeV. The performance for $\Dzero \to {\rm K}^-\pi^+$ measurements was also studied at the lower beam energy of 60 $\GeV/\text{nucleon}$, corresponding to $\sqrtsNN = 10.6$ \GeV.
Charm hadrons were simulated with \pt and rapidity distributions obtained with the {\sc Powheg-Box} event generator~\cite{Alioli:2010xd} for the hard scattering and \pythia~6 for the parton shower and hadronization. 
In the case of collisions at $\sqrtsNN = 16.8$ \GeV, we used $\sigma_{\ccbar}=5$ $\mu$b (based on~\cite{Lourenco:2006vw,Vogt:2001nh} and \powheg). 
This results in a $\Dzero$ yield per event of about 0.006.

The combinatorial background was estimated by simulating pions, kaons, and protons with multiplicity, \pt, and rapidity distributions taken from the parameterisations published by NA49 in Refs.~\cite{Afanasiev:2002mx,Alt:2006dk}. 
Background candidates were built by forming pairs or triplets of reconstructed tracks with proper charge sign combination.
The number of primary particles ($\pi$, $\mathrm{K}$, $\mathrm{p}$) per central \PbPb collision at $\sqrtsNN = 16.8\GeV$ is about 1200, which produce about $3.5\times 10^5$ opposite-sign pairs (i.e.\ \Dzero background candidates) per event, out of which about 8000 have an invariant mass within 0.06\GeVcc from the \Dzero-meson mass.
The signal-to-background ratio for the \Dzero is therefore about $7 \times 10^{-7}$ and needs to be enhanced with the kinematical and geometrical selections.

\begin{figure}[tb]
\begin{center}
\includegraphics[width=0.4\linewidth]{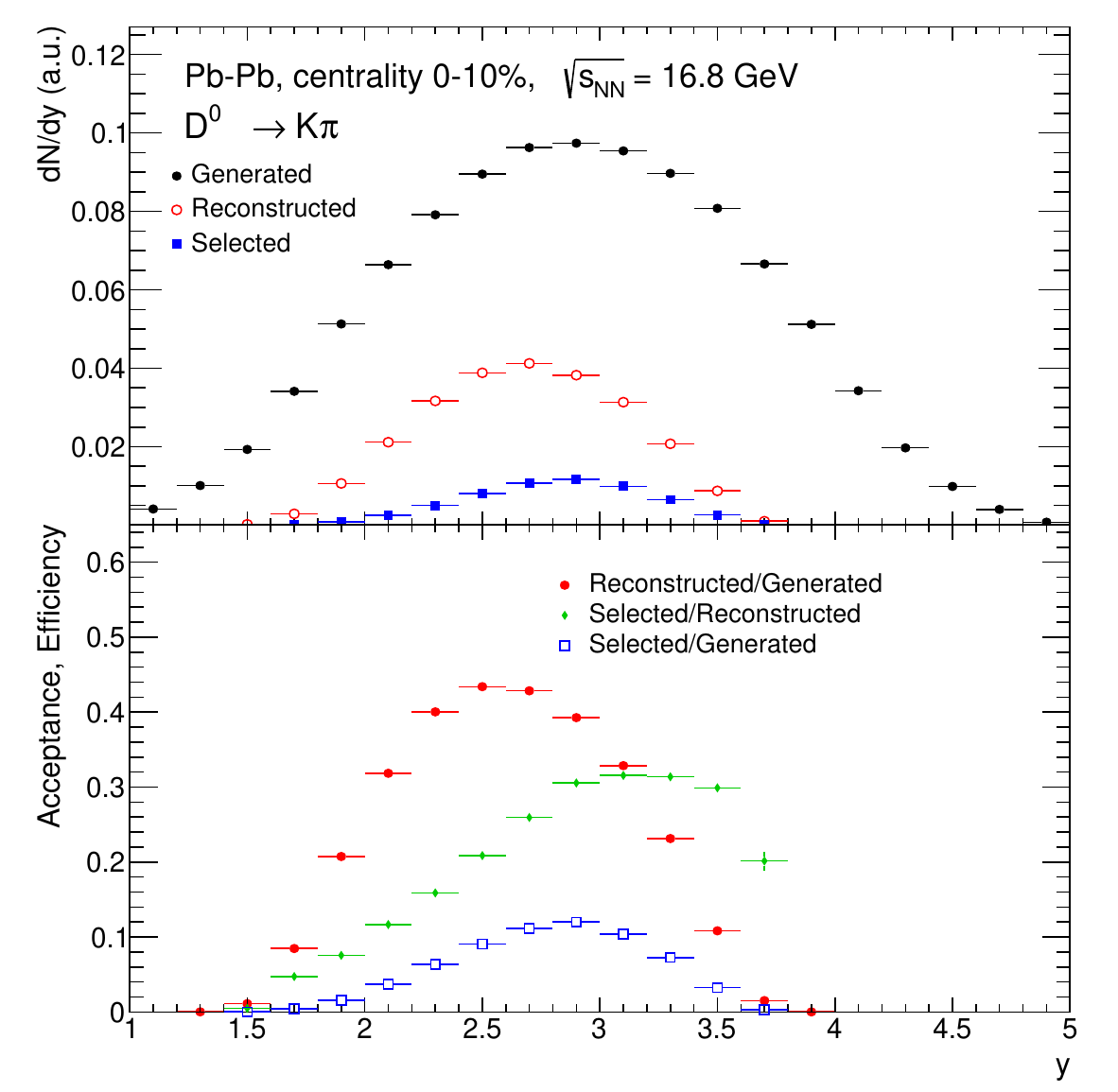}
\includegraphics[width=0.4\linewidth]{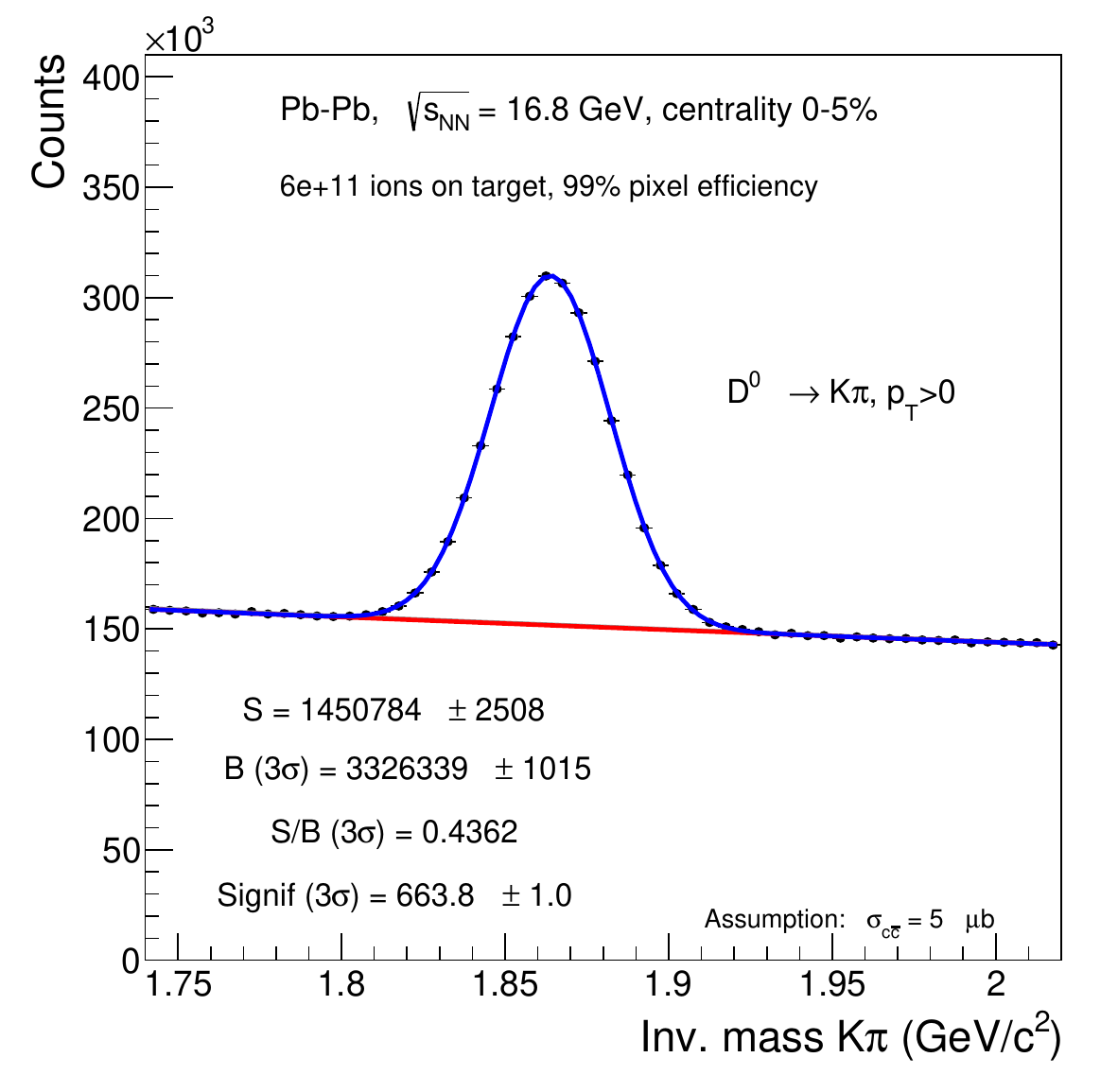}
\caption{Left: Rapidity distributions for $\Dzero \rightarrow \mathrm{K}^- \pi^+$ decays at $\sqrtsNN = 16.8$\GeV at generation level, after reconstruction of decay products in the VT, and after selection cuts based on the decay vertex topology. Right. Projection for the invariant-mass distribution of \Dzero candidates in $5\cdot 10^9$ central \PbPb collisions at beam energy of 150 A GeV.}
\label{fig:D0perf}
\end{center}
\end{figure}

The selection of D mesons was primarily based on the following variables: momentum, $\chi^2$, and impact parameter of the decay tracks, distance of closest approach between the decay tracks, decay length (i.e., distance between primary and secondary vertices), and the cosine of the pointing angle (i.e., the angle between the reconstructed D-meson momentum and the line connecting the primary and secondary vertices).
The rapidity distribution of $\Dzero \rightarrow \mathrm{K}^- \pi^+$ decays that pass the candidate selection criteria adopted in this study is shown in the left panel of Fig.~\ref{fig:D0perf}, along with the rapidity distributions of generated and reconstructed decays. The fraction of candidates with reconstructed decay products passing the selections on the displaced decay vertex topology ranges from about 10\% to about 30\% depending on the D$^0$ rapidity. The right panel of Fig.~\ref{fig:D0perf} presents a projection of the invariant mass distribution of \Dzero candidates in the 5\% most central \PbPb collisions at $\sqrtsNN = 16.8\GeV$, corresponding to a sample of $6 \times 10^{11}$ ions on target, which can be collected in one month of data taking.
It demonstrates that the NA60+/DiCE experiment will enable precise measurements of the \Dzero-meson yield in central \PbPb collisions with a statistical precision well below 1\%, which would allow also for studies in \pt and $y$ intervals and for the determination of the elliptic flow coefficient $v_2$ of D mesons with percent level statistical uncertainty.

Performance studies were also conducted to assess the feasibility of open-charm reconstruction via three-body decay channels of charmed hadrons, which enable measurements of \Dplus and \Ds mesons and of \lambdacplus baryons~\cite{NA60:2022sze}. The outcome of these studies demonstrated that with the NA60+/DiCE setup it will be possible to measure the \Ds meson yield with a significance of approximately 200 in central \PbPb collisions at top SPS energy. Similarly, it will be possible to reconstruct about $3\times 10^4$ \lambdac\ baryons in central \PbPb collisions, achieving a precision of a few percent in statistical uncertainties. This will enable studies of charm quark hadronisation in the SPS energy domain.

\bigskip
\subsection{Charmonia}
\bigskip

No studies of charmonium states below top SPS energy in heavy-ion collisions are available. NA60+/DiCE proposes to measure the J/$\psi$ production in Pb--Pb collisions, in the dimuon decay channel, at the various energies that will be available to the experiment. In the fixed-target domain, cold nuclear-matter effects, which include nuclear shadowing (an initial state mechanism) and break-up of the ${\rm c\overline c}$ pair by the nucleons of the colliding nuclei (a final state mechanism) are known to play a quantitatively important role. Furthermore, there are indications for nuclear break-up cross sections to increase when lowering the collision energy~\cite{NA60:2010wey}. In Pb-Pb collisions, accurate results at top SPS energy were obtained by the NA50/NA60 experiments~\cite{Alessandro:2004ap}, showing a $\sim 30-40\%$ suppression in addition to cold matter effects, which was ascribed to the production of a deconfined medium.

In our experiment, the strategy for the measurement of QGP effects on J/$\psi$ will be based at each energy on the measurement of the nuclear modification factor $R_{\rm AA}$. The yields measured in Pb-Pb collisions need to be normalized to the pp production cross section. The latter will be obtained by means of an extrapolation of results on p-A collisions, collected by exposing to the proton beam various nuclear targets. 
The strength of cold nuclear matter effects will be evaluated by looking at the measured A-dependence of the J/$\psi$ production. A standard procedure, based on the Glauber model, allows an estimate of the expected cold-nuclear matter effects in Pb-Pb collisions, as a function of centrality.

\begin{figure}[ht]
\begin{center}
\includegraphics[width=0.4\linewidth]{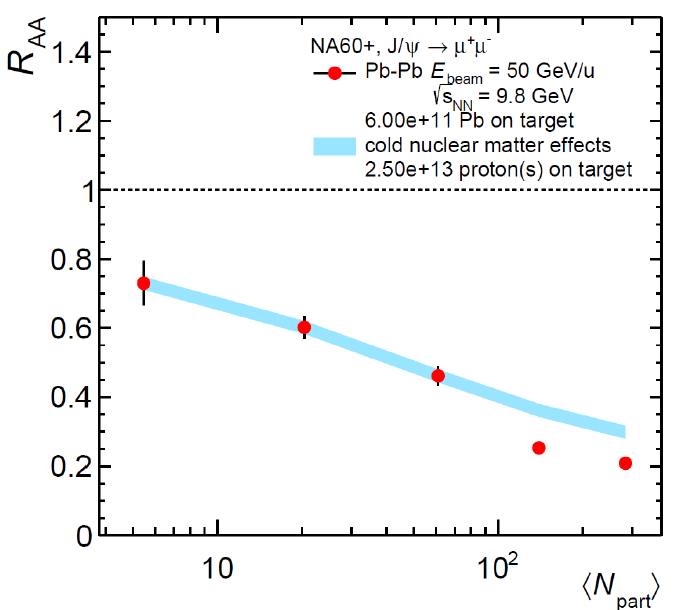}
\includegraphics[width=0.54\linewidth]{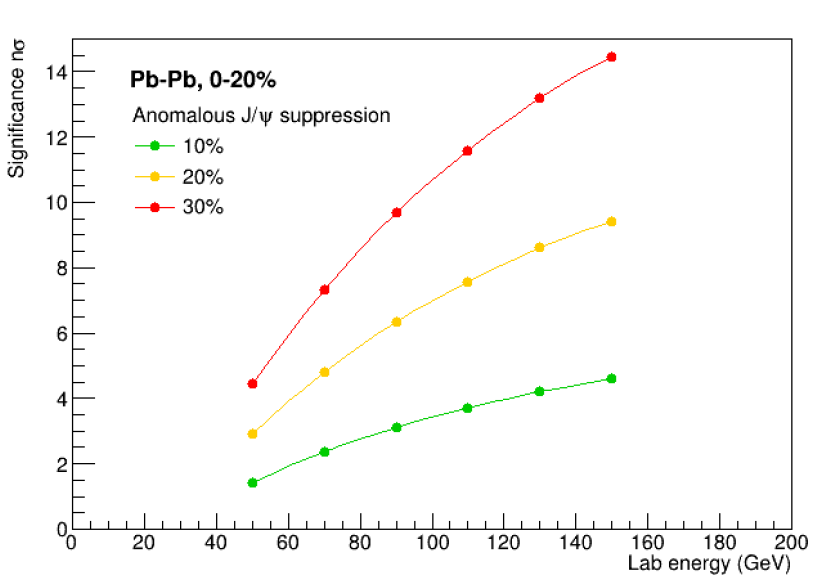}
\caption{(Left) Expected performance for the $R_{\rm AA}$ measurement in Pb-Pb collisions at $E_{\rm lab}=50 A GeV$, as a function of centrality expressed via the number of participant nucleons $N_{\rm part}$ (see text for more details). 
(Right) Significance for the observation of an anomalous suppression on top of the cold nuclear matter effects, as a function of the beam energy. The curves are given for various levels of anomalous suppression ranging from 10 to 30\% and refer to the 20\% more central Pb-Pb collisions.}
\label{fig:jpsi}
\end{center}
\end{figure}

An example of this approach is given in Fig.~\ref{fig:jpsi} (left panel), where the J/$\psi$ $R_{\rm AA}$ corresponding to Pb-Pb collisions at $E_{\rm lab}=50$ GeV is shown. The number of Pb ions and protons on target is indicated in the legend, with the Pb target(s) having a 7.5 mm total thickness (15\% interaction probability) and a 12 mm (Be) + 3 mm (Cu) + 3 mm (Pb) target for p-A collisions. The elementary J/$\psi$ production cross section is introduced starting from parameterization of the (few) available data in this energy range. Cold nuclear matter effects are obtained assuming a break-up cross section of 7.6 mb, a value measured by NA60 in p-A collisions at 158 GeV~\cite{NA60:2010wey}, and their extrapolation to Pb-Pb is shown as the azure line. Finally, a 30\% additional suppression, intended to mimic QGP effects, is added in Pb-Pb collisions for the two more central collision intervals. The comparison between the p-A extrapolation and the Pb-Pb $R_{\rm AA}$ represents a typical plot that is aimed at for this observable.

\bigskip

It has to be stressed that both the strength of cold nuclear matter effects, as well as the level of anomalous suppression, represent an educated guess, as there are no accurate predictions to drive the choice of this quantity. In order to give a more general idea of the potential of the experiment for the J/$\psi$ measurement, we show in Fig.~\ref{fig:jpsi} (right panel) the significance, expressed in terms of number of standard deviations, for the observation of a suppression signal exceeding cold nuclear matter effects by 10, 20 or 30\%. The values are given for the 20\% most central Pb-Pb collisions and are shown as a function of the beam energy per nucleon. As expected, the significance increases with collision energy, due to the larger production cross section. Values larger than $\sim 3\sigma$ can be reached at all energies, for a 20\% anomalous suppression level.

Finally, the detection of the $\psi{\rm (2S)}$ and the $\chi_{\rm c}$ is of great interest, as QGP-related effects are known to depend on the binding energy of the resonance. Their study is more difficult due to their smaller yield. For the $\psi{\rm (2S)}$ dimuon decay, a measurement can realistically be performed dwon to $\sim 120$ A GeV incident energy. The $\chi_{\rm c}\rightarrow {\rm J}/\psi\, \gamma$ decay may also be studied, by detecting the conversion $e^+e^-$ of the photon, but quantitative studies are still in progress.
\section{Timeline and preliminary cost estimate}
\label{sec:timeline}
\bigskip

Our plan is to be able to run the experiment after the end of Long Shutdown 3, in 2029/2030. We foresee to study Pb-Pb collisions at a single center-of-mass energy each year, possibly starting with a 150 AGeV beam. This would allow a rather direct comparison with existing older results from NA60 (thermal dileptons, J/$\psi$), with a larger integrated luminosity. A high-statistics open charm measurement would also become available. Following this first data taking, we aim at exploring  Pb-Pb collisions at five more energies, spanning the SPS coverage down to 20A GeV.
The lower energy point may span over two years, to compensate for the lower production cross sections of the observables that we intend to study. The exact order of the beam energies to be studied will be decided also following the development of the program and the results that will be obtained. This schedule implies that the data taking of the experiment will extend for a few years beyond LS4, down to $\sim 2038$.
If the possibility of using a low-energy primary proton beam will be confirmed, reference data taking at three different energies will be requested inside the period in the year normally devoted to heavy-ion beams. Our preliminary estimates indicate that roughly $3-4$ periods of $\sim 2$ weeks each might be sufficient to collect the needed statistics, of the order of $2.5-5\times 10^{13}$ protons on target, depending on collision energy.

In order to run the experiment with its proposed deadline, we are currently preparing an experiment proposal, to be submitted to the SPSC by mid-2025. If the experiment will be approved, the construction should take place starting in 2026, and be completed in $\sim 3$ years.

The construction cost estimates for the various subsystems were initially provided in the LoI~\cite{NA60:2022sze}, with a range accounting for the number of MAPS engineering runs and different technology options for the muon tracking stations (MWPCs vs. GEMs). Subsequently, MWPCs were demonstrated to offer the best balance of performance and cost. Additionally, NA60+/DiCE will not require dedicated engineering runs for the MAPS. Under these assumptions, the LoI cited a total cost estimate of 10.5 MCHF.
This estimate is presently being reassessed for the experiment proposal. A key revision involves replacing the toroidal magnet (originally estimated at 3.8 MCHF) with the MNP33 dipole. This substitution is expected to generate significant cost savings, even when accounting for additional expenses not included in the original estimate: (i) costs associated with relocating, refurbishing, and powering existing magnets, and (ii) required modifications and upgrades to both the H8 beam line and experimental hall infrastructure.
%
%
%

%
\newpage
\bibliographystyle{utphys}
\bibliography{main}

\newpage
\appendix
\section*{Appendix: NA60+/DiCE Collaboration}
\label{app:collab}
\bigskip
\begingroup
\begin{flushleft}
C.~Ahdida\Irefn{cern}\And
G.~Alocco\Irefnn{utorino}{torino}\And
M.~Arba\Irefn{cagliari}\And
R.~Arnaldi\Irefn{torino}\And
S.~Beol\`e\Irefnn{utorino}{torino}\And
A.~Beraudo\Irefn{torino}\And
J.~Bernhard\Irefn{cern}\And
L.~Bianchi\Irefnn{utorino}{torino}\And
E.~Borisova\Irefn{weizmann}\And 
S.~Bressler\Irefn{weizmann}\And 
S.~Bufalino\Irefnn{disat}{torino}\And
R.~Cerri\Irefnn{utorino}{torino}\And
C.~Cical\`o\Irefn{cagliari}\And
S.~Coli\Irefn{torino}\And
P.~Cortese\Irefnn{piemonte}{torino}\And
A.~Dainese\Irefn{padova}\And
H.~Danielsson\Irefn{cern}\And
J.~Datta\Irefn{stonybrook}
A.~De Falco\Irefnn{ucagliari}{cagliari}\And
A.~Drees\Irefn{stonybrook}\And
L.~Epshteyn\Irefn{weizmann} \And 	
A.~Ferretti\Irefnn{utorino}{torino}\And
F.~Fionda\Irefnn{ucagliari}{cagliari}\And
M.~Gagliardi\Irefnn{utorino}{torino}\And
G.M.~Galimberti\Irefn{cagliari}\And
G.~Jin\Irefn{hefei}\And
F.~Geurts\Irefn{rice}\And
V.~Greco\Irefnn{ucatania}{catania}\And
L.~Hu\Irefn{hefei}\And
L.~Levinson\Irefn{weizmann}	\And 
F.~Li\Irefn{hefei} \And
W.~Li\Irefn{rice}\And
Z.~Liu\Irefn{hefei} \And
D.~Marras\Irefn{cagliari}\And
M.~Masera\Irefnn{utorino}{torino}\And
A.~Masoni\Irefn{cagliari}\And
F.~Mazzaschi\Irefn{cern}\And
P.~Mereu\Irefn{torino}\And
L.~Micheletti\Irefn{torino}\And
A.~Milov\Irefn{weizmann}\And 
L.~Mirasola\Irefnn{ucagliari}{cagliari}\And
A.~Mulliri\Irefnn{ucagliari}{cagliari}\And
L.~Musa\Irefn{cern}\And 
C.~Oppedisano\Irefn{torino}\And
N.~Pacifico\Irefn{cern}\And
M.~Pennisi\Irefnn{utorino}{torino}\And
S.~Plumari\Irefn{ucatania}\And
T.~Prebibaj\Irefn{cern}\And
F.~Prino\Irefn{torino}\And
M.~Puccio\Irefn{cern}\And
C.~Puggioni\Irefn{cagliari}\And
R.~Rapp\Irefn{tamu}\And
A.~Rossi\Irefn{padova}\And
V.~Sarritzu\Irefnn{ucagliari}{cagliari}\And
B.~Schmidt\Irefn{cern}\And
E.~Scomparin\Irefn{torino}\And
D.~Sekihata\Irefn{tokyo}\And
Q.~Shou\Irefn{fudan}\And
S.~Siddhanta\Irefn{cagliari}\And
R.~Shahoyan\Irefn{cern}\And
X.~Su\Irefn{hefei}\And
Z.~Tang\Irefn{hefei}\And
S.~Trogolo\Irefnn{utorino}{torino}\And
M.~Tuveri\Irefn{cagliari}\And
A.~Uras\Irefn{lyon}\And
G.~Usai\Irefnn{ucagliari}{cagliari}\And
M.~Van~Dijk\Irefn{cern}\And
E.~Vercellin\Irefnn{utorino}{torino}\And
I.~Vorobyev\Irefn{cern} \And
D.~Zavazieva\Irefnn{weizmann}{bgu} 
\renewcommand\labelenumi{\textsuperscript{\theenumi}~}
\end{flushleft}

\bigskip
\renewcommand\theenumi{\arabic{enumi}~}
\begin{Authlist}
\item \Idef{cern}European Organization for Nuclear Research (CERN), Geneva, Switzerland
\item \Idef{utorino}Dipartimento di Fisica dell Universit\`{a} di Torino, Turin, Italy
\item \Idef{torino}INFN, Sezione di Torino, Turin, Italy
\item \Idef{cagliari}INFN, Sezione di Cagliari, Cagliari, Italy
\item \Idef{weizmann}{Department of Particle Physics and Astrophysics, Weizmann Insitute of Science, Rehovot, Israel}
\item \Idef{disat}Dipartimento DISAT del Politecnico di Torino, Turin, Italy
\item \Idef{piemonte}Dipartimento di Scienze e Innovazione Tecnologica dell'Universit\`{a} del Piemonte Orientale, \\Alessandria, Italy
\item \Idef{padova}INFN, Sezione di Padova, Padova, Italy
\item \Idef{stonybrook}Department of Physics and Astronomy, Stony Brook University, SUNY, Stony Brook, New York, USA
\item \Idef{ucagliari}Dipartimento di Fisica dell'Universit\`{a} di Cagliari, Cagliari, Italy
\item \Idef{hefei} Department of Modern Physics, University of Science and Technology of China, Hefei, Anhui, China
\item \Idef{rice}Department of Physics and Astronomy, Rice University, Houston, Texas, USA
\item \Idef{ucatania}Dipartimento di Fisica e Astronomia dell'Universit\`{a} di Catania, Catania, Italy
\item \Idef{catania}INFN, Laboratori Nazionali del Sud, Catania, Italy
\item \Idef{tamu}Cyclotron Institute and Department of Physics and Astronomy, Texas A\&M University, College Station, Texas, USA
\item \Idef{tokyo}University of Tokyo, Tokyo, Japan
\item \Idef{fudan}Institute of Modern Physics, Fudan University, Shanghai, China
\item \Idef{lyon}Institut de Physique des 2 Infinis de Lyon, Université de Lyon, CNRS/IN2P3, Lyon, France
\item \Idef{bgu}{Faculty of Engineering Sciences, Ben Gurion University of the Negev, Beer Sheva, Israel}

\end{Authlist}
\endgroup

\end{document}